\documentclass[runningheads]{llncs}
\usepackage{graphicx}
\usepackage{verbatim}
\usepackage{listings}
\usepackage{xcolor}
\usepackage{wrapfig}
\usepackage{subcaption}
\usepackage{hyperref}
\usepackage{cite}
\usepackage{fnpct}

\definecolor{mGray}{rgb}{0.5,0.5,0.5}
\begin{document}
\title{Implementing the Comparison-Based External Sort}
%
%
\author{Michael Polyntsov\inst{1, 2}\orcidID{0000-0001-7356-2504} \and
Valentin Grigorev\inst{2}\orcidID{0000-0003-4235-3712} \and
Kirill Smirnov\inst{1, 2}\orcidID{0000-0003-4727-3455}
\and
George Chernishev\inst{1, 2}\orcidID{0000-0002-4265-9642}}
\authorrunning{M. Polyntsov et al.}
%
\institute{ 
Saint-Petersburg State University, Russia
\and 
PosDB Team\\
\email{\{polyntsov.m,valentin.d.grigorev,kirill.k.smirnov,chernishev\}@gmail.com}}
\maketitle              

\begin{abstract}

In the age of big data, sorting is an indispensable operation for DBMSes and similar  systems. Having data sorted can help produce query plans with significantly lower run times. It also can provide other benefits like having non-blocking operators which will produce data steadily (without bursts), or operators with reduced memory footprint.

Sorting may be required on any step of query processing, i.e., be it source data or intermediate results. At the same time, the data to be sorted may not fit into main memory. In this case, an external sort operator, which writes intermediate results to disk, should be used. 

In this paper we consider an external sort operator of the comparison-based sort type. We discuss its implementation and describe related design decisions. Our aim is to study the impact on performance of a data structure used on the merge step. For this, we have experimentally evaluated three data structures implemented inside a DBMS.  

Results have shown that it is worthwhile to make an effort to implement an efficient data structure for run merging, even on modern commodity computers which are usually disk-bound. Moreover, we demonstrated that using a loser tree is a more efficient approach than both the naive approach and the heap-based one.

\end{abstract}

\section{Introduction}

\footnotetext{This paper was accepted to MegaData@ADBIS'22.}

Sorting is a very important operation in any data processing system. For example, in DBMSes leveraging sorted data may allow efficient implementation of operators~\cite{10.5555/1972515} such as sort-merge join, which is one of the most popular approaches to join two tables larger than available memory in a reasonable time. Another example is performing aggregation over data sorted on aggregation attributes. In this case, it is possible to implement a non-blocking aggregation operator, i.e., an operator which can start producing results without having to read all input records.

There are two classes of sorting operators~--- internal (in-memory) and external (disk-based) sorts. The former assumes that all data fits into main memory and the latter has to write intermediates to disk. Recent research mostly focuses on an in-memory type because of a surge of interest that in-memory processing currently experiences~\cite{10.1145/3448016.3457319,8855628,8638394}. Such systems try to avoid external sorting, since it significantly degrades performance. Despite all this, disk-based systems are still in use\footnote{As of July 2022, disk-based systems PostgreSQL and MySQL were in the top-5 most used according to DB-Engines Ranking (\url{https://db-engines.com/en/ranking}).} and research in this area is relevant.

Thus, efficient implementation of the sorting operator is crucial for the performance of a DBMS as a whole. Such implementation heavily depends on the amount of an engineering effort put into the source code, which is specific for each particular system. For example, it may include adjusting implementation to various system parameters (e.g. disk block size, cache line size, etc.) and writing efficient code in general. However, some high-level techniques can be reused between systems, and thus they are of scientific interest. For example, such techniques as key normalization or record surrogate usage are discussed in Graefe's survey~\cite{GraefeSortingInDBMS}.

In this paper we study the implementation of efficient external sorting in a disk-based column-store. Aside from the aforementioned application, in column-stores, external sort has another important one: it addresses the out-of-order probing problem~\cite{Abadi:2013:DIM:2602024}. The particular research focus of this paper is the impact of data structure choice on the performance of the comparison-based external sort. This data structure is used during run merging and allows to reduce the number of comparisons. More specifically we pose two following research questions: 1) is it beneficial to use a specialized data structure on the merge step in a contemporary environment (commodity PCs), and 2) which kind is the best.

To answer them, we have implemented an external sort operator inside PosDB~\cite{Chernishev:2018:PAO,DBLP:conf/dolap/ChernishevGGK022}, a distributed column-store engine. In the operator's core, we have devised three approaches to implementing the merge: naive (comparing all to all), a loser tree, and a binary heap.

To validate the quality of our implementation we first compare it with PostgreSQL's external sort. Then we evaluate these three data structures by comparing them with each other using a synthetic and a real dataset. 

The paper is organized as follows. In Section~\ref{sec:background} we provide a concise introduction into external sorting. Then, in Section~\ref{sec:relwork} we discuss the related work and provide a motivation for our study. Next, in Section~\ref{sec:solution} we describe the proposed approaches and their implementation. In Section~\ref{sec:experiments} we explain conducted experiments and their results. Finally, we conclude this paper with Section~\ref{sec:conc}.

\section{Background}\label{sec:background}
Sorting, because of its importance, has been studied very extensively. Sorting algorithms fall into two categories: external and in-memory (or internal~\cite{KnuthSort}) sorting. In this paper we will focus on external sorting. An external sorting algorithm can be either comparison-based or partition-based. Algorithms of the partition-based class distribute input tuples into buckets using several pivot values (selected by the algorithm) while ensuring the following properties:
\begin{itemize}
    \item each bucket covers some data range,
    \item all data ranges do not overlap,
    \item a union of all data ranges covers the whole range.
\end{itemize}
Then the algorithm sorts the contents of individual buckets and in doing so it is ensured that each bucket fits into memory. The result is an ordered sequence of buckets, each containing a sorted range.

The partition-based sorting algorithm has two degrees of freedom: the algorithm that sorts the buckets in memory and pivot selection.

Comparison-based algorithms generate sorted runs (i.e., in this case, data ranges intersect) and then merge them into one long run. They consist of two stages: run generation and run merging. These stages can be either completely independent and consecutive, or alternate (known as oscillating sort~\cite{KnuthSort}). Consequently, there are two approaches to run generation:

\begin{enumerate}
    \item Read tuples from disk for as long as there is enough available memory; sort these tuples using some internal sorting algorithm;
    \item The algorithm maintains a data structure in memory with a priority queue interface. On each iteration the lowest value from the data structure is retrieved and moved to a buffer or written to the disk. Then algorithm inserts the next value from the input data into the data structure. If the next value is smaller than the last retrieved, it should be marked as belonging to the next run. When the data structure contains only such marked values, the current run is treated as completed. After this, the new one is started. This method is known as the replacement selection~\cite{KnuthSort}. The tournament tree~\cite{KnuthSort} can be used to implement this method.
\end{enumerate}

The second approach was more popular in the past. On average, it generates runs that are twice the size of available memory, which can be useful in a memory-constrained environment. Nowadays, available RAM sizes have grown significantly and the Quicksort became more popular. 

In the past, tapes were used instead of disks for persistent storage. Of course, there can be more runs than tapes. Different algorithms exist to merge runs and distribute them over tapes, e.g. multiway merge, polyphase merge, cascade merge, etc. These algorithms are called merge patterns. Disks are the ubiquitous storage type nowadays, but the concept of tape as an abstraction can still be useful in order to save disk space by reusing input tapes as output tapes. This technique is still used in industrial systems\footnote{\url{https://github.com/postgres/postgres/blob/REL_14_STABLE/src/backend/utils/sort/tuplesort.c}}.

In addition to the merge pattern, comparison-based sorting algorithms differ in the algorithm for selecting the smallest value among the first elements of the merged runs. The naive approach, and therefore the asymptotically slowest, is to check all candidates. A more efficient approach here is to use a data structure that speeds up finding the smallest element. For this purpose, a heap or a tournament tree (either loser or winner) are the most common choices.


\section{Related Work and Motivation}\label{sec:relwork}

\subsection{Related Work}
Over the years, there has been a lot of research on external sorting. The algorithmic theoretical aspects of external comparison-based sorting are very well understood. That is, in-memory sorting algorithms, run merge algorithms and the appropriate data structures for single-processor environment are known and have not largely changed over time since Knuth's survey~\cite{KnuthSort}. Most of the current research in non-distributed external sorting is focused on efficient implementation of an algorithm on modern hardware, i.e., implementing sorting that fully utilizes all available hardware capabilities. The two most popular concerns are how to utilize caches efficiently and how to mitigate disk bandwidth limitations. More generally, one can classify the most of the current research papers on external sorting as follows:

\begin{itemize}
    \item techniques to improve sorting performance by using hardware resources more efficiently, namely:
    \begin{itemize}
        \item CPU and caches
        \item HDD/SSD 
        \item GPU
    \end{itemize}
    \item novel external sorting algorithms that make heavy use of parallelism;
    \item distributed external sorting algorithms. 
\end{itemize}

In the paper~\cite{AlphaSort} authors describe the AlphaSort algorithm. It was designed with memory hierarchy in mind and therefore exhibits cache-friendly behavior. The Quicksort is used for in-memory sorting since: 1) it has better cache locality and is less CPU-intensive than the replacement-selection, and 2) modern computers have enough RAM to produce sufficiently long runs. Runs are merged using a tournament tree. The authors compare different approaches to storing tuples during sorting and merging. Possible alternatives are: sort tuples itself, sort pointers to tuples, sort pairs of key and pointer, sort pairs of key prefix and pointer. The authors conclude that sorting the tuples itself can be faster if tuples are short enough. Otherwise, sorting the tuple-pointer and key-prefix pairs is most efficient. A parallel version of the algorithm for multiprocessor systems is also presented.

The authors of the study~\cite{AdaptiveSort} focus on external sort operator in memory-inconsistent environment. They consider real-time or goal-oriented DBMSes whe\-re the available memory for a query (or transaction) may change on-the-fly due to other concurrent transactions. It may either be shrunk since some higher priority transaction requires more memory, or increased because some transaction has finished running and freed its memory. The authors propose an approach called dynamic splitting to efficiently address fluctuations that occurred during the merge phase. They compare dynamic splitting with two other known approaches: suspension and paging, which can only be used in case when the memory is reduced. The authors conclude that using replacement selection with block writes at the run generation stage and dynamic splitting at the run merge stage is the most efficient way to handle memory fluctuations.

In the paper~\cite{Leyenda} the authors present the Leyenda algorithm. It is a novel parallel external sorting algorithm with state-of-the-art performance. Hybrid approach is proposed to generate runs, namely an adaptive parallel most-significant-digit Radix sort. To address the skewed input data issue, Leyenda load balances small buckets between threads and splits big buckets to process them using multiple threads. If the size of a bucket is less than a certain threshold, the Quicksort algorithm is used to sort it. K-way-merge algorithm is used to merge runs. In order to achieve efficient I/O Leyenda leverages parallelism and memory mapping. \texttt{mmap()} is used to map disk memory to OS pages, allowing parallel I/O operations to be performed on it. Leyenda places most frequently accessed data to the caches during both Radix sort and I/O operations.

Paper~\cite{DeWittSplitting} is focused on the splitting stage of a parallel multiprocessor external sort in shared-nothing environment. It is assumed that the sorting algorithm works in a Samplesort~\cite{Samplesort} fashion: first, input tuples are split into buckets which are distributed among the processors so that the $i$-th processor has the bucket with tuples that are less than the ones on the $(i+1)$-th processor. Next, the tuples are sorted locally on each processor, and then, sorted buckets are concatenated. The study compares two approaches to splitting input tuples at the first stage: exact splitting~\cite{ExactSplitting} and probabilistic splitting. The authors discuss implementations and possible trade-offs. Based on the conducted experiments, they conclude that the probabilistic splitting is more efficient and, with the growth of the input file size the performance gap only increases.

The authors of the study~\cite{GPUTeraSort} are focused on sorting on GPUs: they propose a novel external sorting algorithm GPUTeraSort. The idea is that all computationally-intensive and memory-intensive operations are handled by the GPU while the CPU manages resources and I/O operations. GPUTeraSort has five stages that are pipelined using independent threads: Reader, Key-Generator, Sorter, Reorder, and Writer. The algorithm sorts key-pointer pairs using an improved bitonic sort on the GPU. Disk striping is used to achieve peak I/O performance.

The authors of the paper~\cite{SpeedingUpExternalMergesort} describe various techniques for speeding up I/O operations of a comparison-based external sort on HDD. They review and compare two layout strategies: contiguous layout and interleaved layout. The former is straightforward (and most widely used) based on the idea of storing each run continuously. This implies sequential writing in the run generation phase, but results in random accesses during the merge phase. The latter stores blocks from different runs in an interleaved manner to reduce disk seeks during the merge phase. The authors compare three reading strategies: forecasting, double buffering and planning strategy. They conclude that a contiguous layout and the planning read strategy are the most efficient. 

Study~\cite{FlashMemorySorting} is focused on external sorting on a system with NAND flash memory as a secondary storage (SSD). Authors review NAND flash memory properties that are important for the external sorting algorithm and propose techniques to effectively leverage these properties. Unclustered sorting is a sorting algorithm that first sorts key-pointer pairs (named tags) and then rearranges input tuples using sorted tags. Whereas clustered sorting algorithm sorts tuples straightaway. Unclustered sorting is designed to reduce the number of disk reads and writes, which has direct impact on SSD performance. For the same purpose, the concept of reused pages is proposed. Authors state that clustered sorting can be more efficient if input tuples are small enough. A decision rule to select the more suitable algorithm at run time is presented.


In the paper~\cite{SortingOnNetworks} authors propose external sorting algorithms on Network of Workstations, measure their performance and compare it to performance of algorithms on shared-memory computers. They discuss different approaches to perform I/O operations in order to achieve best performance. The authors compare different algorithms for in-memory sorting and identify the most efficient one based on the conducted experiments. Four sorting algorithms on Network of Workstations are proposed and discussed: One-Pass Single-Node, One-Pass Parallel, Two-Pass Single-Node, Two-Pass Parallel. Comparison-based algorithms are used for external sorting. Also, disk striping and overlapping of computation and I/O are discussed in-depth.

In the study~\cite{LoadBalancedSorting} the authors are focused on external sorting on the shared-nothing architecture. They propose a novel parallel load-balanced multiple-input multiple-output algorithm. It implements multiprocessor multi-way merge that guarantees load balancing of execution, communication, and output. That is, all processors are loaded evenly during sorting, no network congestion occurs and, after sorting is finished, there is no redistribution skew. The authors compare their algorithm with external sorting algorithms based on the exact and probabilistic splitting. They conclude that their approach is more efficient than exact splitting and comparable to probabilistic splitting.

\subsection{Motivation}

Thus, to the best of our knowledge, there are no studies that consider the design of a comparison-based sort and evaluate the choice of the data structure used on the merge step for implementing the external sort operator in DBMSes. 

At the same time, it is of interest to industry and leads to a number of important questions. It is a well-known fact that commodity computers with a single disk are I/O-bound and thus, there may be no point to invest effort into perfecting the in-memory part of the algorithm. Next, if it is nevertheless beneficial, further questions arise~--- which data structure is the best and how does varying parameters affect it.



\begin{figure}[h!]
    \centering
    \includegraphics[width=0.9\linewidth]{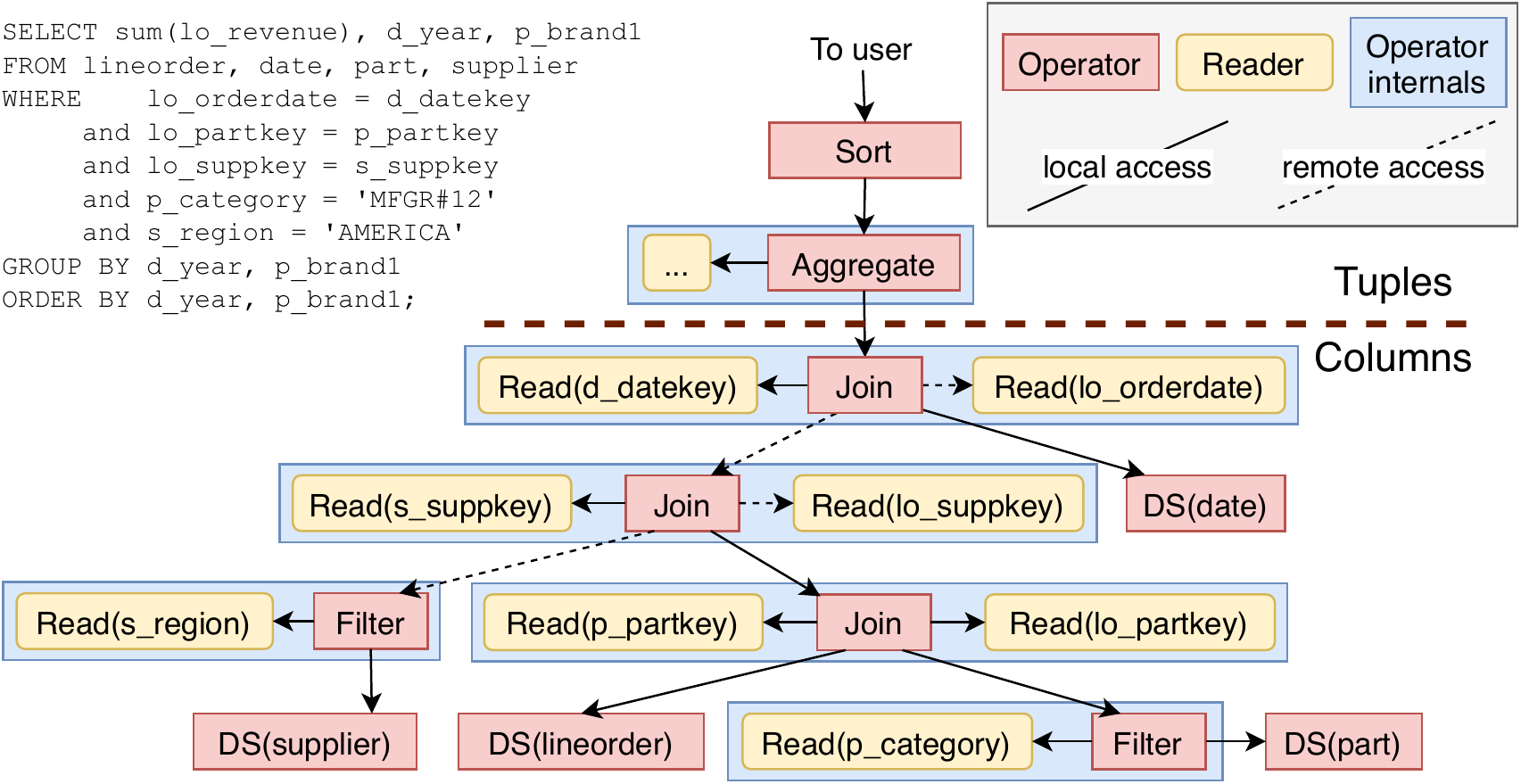}
    \caption{Ultra-late materialization plan example}
    \label{fig:plan-example}
\end{figure}

\section{Proposed Solution}\label{sec:solution}
\subsection{PosDB Basics}
To experimentally evaluate the proposed solution, we used PosDB~\cite{Chernishev:2018:PAO,DBLP:conf/dolap/ChernishevGGK022}, a column-oriented parallel distributed disk-based query engine. PosDB uses the Volcano model~\cite{GraefeVolcano} to represent query execution plans. That is, each plan is an operator tree with the edges describing data flows. Each operator has an iterator interface. PosDB implements block-oriented processing (operators exchange blocks of tuples), which is more efficient~\cite{BlockOriented} than tuple oriented processing. Since PosDB is column-oriented, i.e., stores tables in a columnar format, at some point during the query execution, it is necessary to convert the column representation to a tuple-based one. This moment is called the materialization point, and the stage of query execution where this conversion takes place is determined by the materialization strategy. In the query plan presented on Fig.~\ref{fig:plan-example} materialization point is indicated by a brown dotted line. PosDB supports several materialization strategies, namely: early, late and ultra-late materialization, and partially supports hybrid materialization. Early materialization strategy is the most common one in contemporary column-stores, its idea is to perform materialization before (or sometimes in) filter operators. Late materialization, by contrast, tries to postpone materialization for as long as possible (until first or second join, depending on system). Ultra-late materialization is a PosDB refinement for late materialization described in~\cite{DBLP:conf/dolap/ChernishevGGK022}. Finally, hybrid materialization is a strategy where positions are processed simultaneously with the tuple values. 

\begin{figure}[h!]
    \centering
    \includegraphics[scale=0.6]{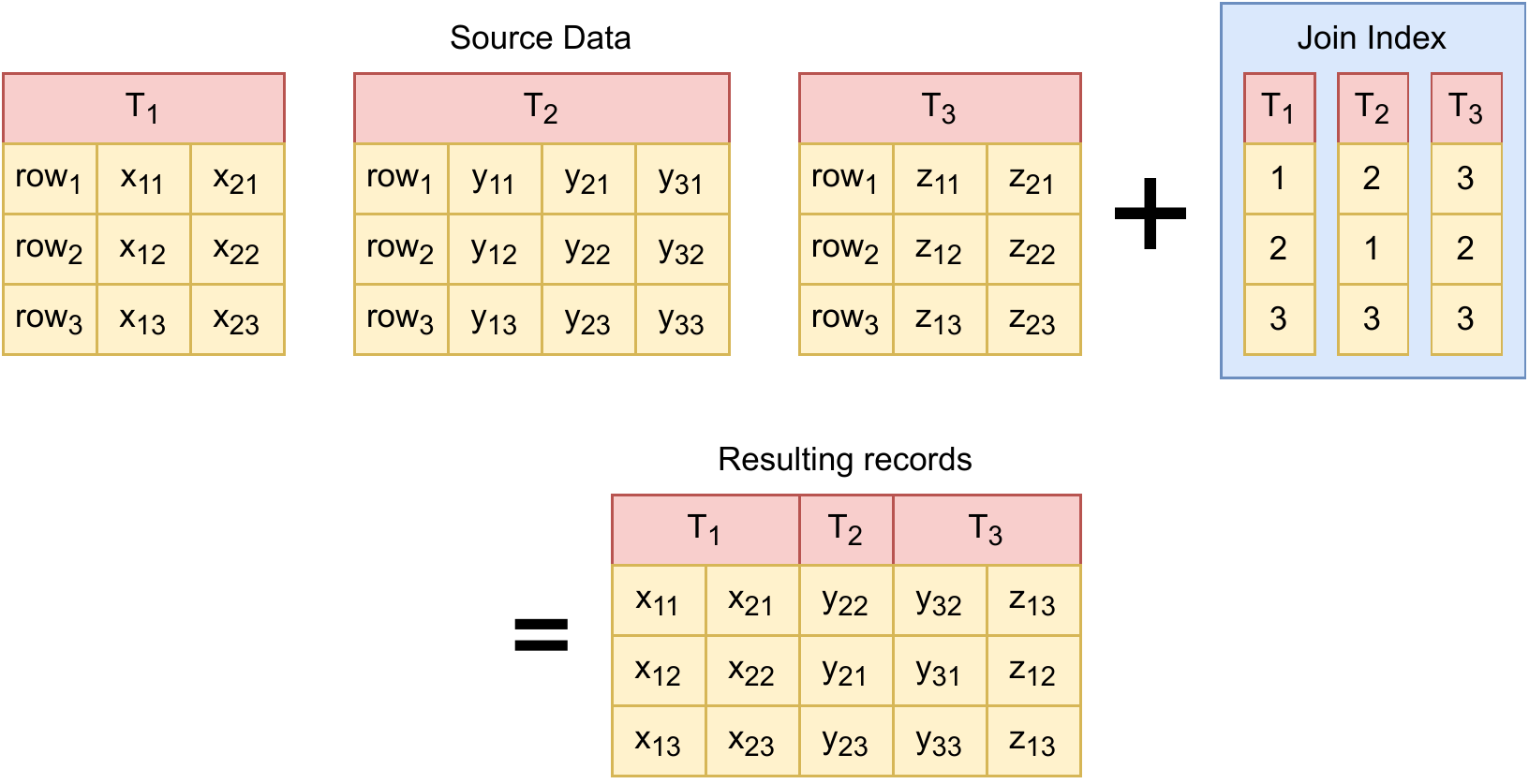}
    \caption{Example of join index and its usage}
    \label{fig:ji-example}
\end{figure}

To handle all of these strategies, different types of operators are implemented in PosDB: position-based, tuple-based, and hybrid operators. Position-based operators use the generalized join index~\cite{JoinIndex} (an example is shown in Fig.~\ref{fig:ji-example}) to effectively represent positional data. The join index only contains information about the positions, not the values themselves, which is not sufficient to perform some operations, such as joining or filtering. The functionality to retrieve values via positions in PosDB is provided by auxiliary entities called readers~\cite{Chernishev:2018:PAO}. There are several types of readers, for example, ColumnReader, PartitionReader, SyncReader. ColumnReader retrieves values of a specified attribute locally, PartitionReader retrieves values of a specified partition either locally or remotely. SyncReader is a composite reader that manages several simpler ones to extract the values of several attributes synchronously. Readers, in turn, use access methods~\cite{Hellerstein} to provide their functionality.

Thus, PosDB is able to emulate row-based systems by exploiting early materialization. Such emulation was used to experimentally evaluate the implemented external sorting algorithms. 

\subsection{Operator Implementation}
External sort is implemented as a separate operator in the Volcano model. There are several external sort operators in PosDB:
\begin{enumerate}
    \item \textbf{Position-based value sort}. In this case, the operator accepts join index as input and has to obtain values since it needs them for sorting.
    \item \textbf{Position-based position sort}. Input is the same, but in this case the operator does not need values, since it sorts positions.
    \item \textbf{Tuple-based}. This is the classic sort, where operator accepts blocks of tuple data and sorts them. Since in PosDB tuple-based representation lacks positions, it is a value sort.
\end{enumerate}

Position-based value sort operator may be beneficial for late materialization (and therefore interesting for further research), since it sorts key-position pairs similar to unclustered (tag) sorting, which is stated~\cite{FlashMemorySorting} to be more efficient on SSD if tuples are big enough. However, it is beyond the scope of this paper. 

The same is true regarding the position-based position sort. Sorting positions is essential for addressing the out-of-order probing problem inherent for column-stores~\cite{Abadi:2013:DIM:2602024}. 

The focus of this paper is the latter, classic variant of external sorting~--- the tuple-based operator. It implements a comparison-based external sorting algorithm on tapes with the Introsort~\cite{Introsort} for run generation and polyphase merge~\cite{KnuthSort} for run merging. The algorithm works as follows:

\begin{enumerate}
    \item Calls \texttt{GetNext()} of child operator until all available memory is filled;
    \item Sorts collected tuples in-memory using \texttt{std::sort} implemented as Introsort;
    \item Using the polyphase merge pattern, selects the abstract tape to which the current run will be flushed;
    \item Flushes current run to the selected tape;
    \item Repeats steps 1--4 until input is fully processed;
    \item Merges runs from the tapes according to the polyphase merge pattern until there is only one run left.
\end{enumerate}

The implemented external sort tuple-based operator sorts pointers to tuples, not tuples themselves. 
Even though run merging can be efficiently performed in a single pass in many simple queries, multilevel merge may still may be necessary to efficiently execute a complex query plan, e.g. a plan with several sort and hash operations~\cite{GraefeSortingInDBMS}. Taking this into account, it was decided to implement abstract tapes to store runs. Run merging (and distributing over tapes) is performed using polyphase merge, which, however, falls back to ordinary multiway merge when there are more tapes than runs. Abstract tapes can also be used to efficiently reuse disk space during the merge stage, as is done in PostgreSQL\footnote{See footnote 4.}.
External sort operator completely bypasses PosDB buffer manager and writes generated runs directly to the disk, it is essential~\cite{GraefeSortingInDBMS} for optimized I/O. Introsort was chosen for the in-memory sorting algorithm for the same reasons that make Quicksort superior~\cite{AlphaSort} to the replacement selection, namely:

\begin{enumerate}
    \item Quicksort results in far fewer cache misses, which is very important for the performance of modern CPUs;
    \item Quicksort is less CPU intensive;
    \item With today's main memory sizes, there is no need for runs to be twice the amount of available memory (this is what made the replacement selection very attractive in the past);
    \item With Quicksort (or any other in-place sorting algorithm) it is possible to copy each record only once during run generation stage, while replacement selection needs at least two copy operations~\cite{GraefeSortingInDBMS}.
\end{enumerate}

For all of these reasons, replacement selection is widely considered obsolete, and PostgreSQL abandoned it in favor of the Quicksort algorithm a few years ago. Also, none of top five solutions in the ACM SIGMOD 2019\footnote{\url{http://sigmod19contest.itu.dk/}} Programming Contest (in which external sorting was one of the challenges) used it. Among them, two solutions implemented some kind of radix sort for in-memory sorting, other two implemented Quicksort with various optimizations and Leyenda implemented a hybrid approach. Despite the fact that the Radix sort is the most efficient sorting algorithm for some specific data types (e.g. integers), it is difficult to implement it as a general-purpose algorithm for sorting arbitrary data.

During the run merging stage, it is necessary to select the smallest element among the frontmost tuples of runs being merged. It can be done using various approaches, and we implemented and experimentally evaluated three, namely: 1) naive, 2) using the tree of losers data structure, and 3) using a binary heap. All disk operations are buffered via basic \texttt{std::fstream} buffers.

\textbf{Naive.} The naive approach is straightforward and does not use an auxiliary data structure. It simply iterates over first tuples of input runs and selects the smallest ones. The selected tuple is then flushed to the output tape and replaced with the next tuple from its corresponding run.

\textbf{Binary heap.} This approach maintains a binary heap in memory to speed up finding the smallest element. First, it constructs the binary heap and fills it with pointers to the first tuples of runs. Second, it removes the root from the heap and flushes tuple pointed to by the root to the output tape. It then reads the next tuple from the run of the previously removed tuple and inserts a pointer to it to the heap.

\textbf{Loser tree.} This is the same approach as with a binary heap, but instead of a heap, a loser tree~\cite{KnuthSort} data structure is used. 

\section{Experiments}\label{sec:experiments}

To evaluate and compare the performance of the implemented algorithms in a DBMS environment, we have performed the following experiments:

\begin{enumerate}
    \item Comparison with PostgreSQL in order to ``validate'' the quality and correctness of our implementation;
    \item Evaluation of the three approaches on the synthetic Star Schema Benchmark~\cite{SSB};
    \item Evaluation of the three approaches on a real dataset, namely, TripData~\cite{TripData}.
\end{enumerate}

Experiments were performed using the following hardware and software configuration:

\begin{itemize}
    \item Hardware: Intel Core i5-7200U CPU @ 2.50GHz × 4, 8 GiB RAM, 240GB KINGSTON
SA400S3;
    \item Software: Ubuntu 20.04.4 LTS x86\_64, Kernel 5.13.0-40-generic,  gcc 9.4.0, PostgreSQL 12.10.
\end{itemize}

For the first two experiments, we decided to use a simple query with an \texttt{ORDER BY} clause over the \texttt{lineorder} table from the Star Schema Benchmark. SSB was used with a scale factor of 35 (over 10 GBs). In this experiment the table was modified by excluding columns that do not participate in any query that comes with this benchmark.

The query under consideration over \texttt{lineorder} was as follows:

\begin{lstlisting}[language=sql, numberstyle=\color{mGray}, label={lst:query1}, numbers=left, basicstyle=\small, xleftmargin=2em,frame=single,framexleftmargin=1.5em]
SELECT
  LO_CUSTKEY, LO_DISCOUNT, LO_EXTENDEDPRICE, LO_ORDERDATE,
  LO_ORDERPRIORITY, LO_ORDTOTALPRICE, LO_PARTKEY,
  LO_QUANTITY, LO_REVENUE, LO_SUPPKEY, LO_SUPPLYCOST
FROM LINEORDER
ORDER BY LO_ORDERDATE;
\end{lstlisting}

\begin{figure}[h!]
    \centering
    \includegraphics[scale=0.45]{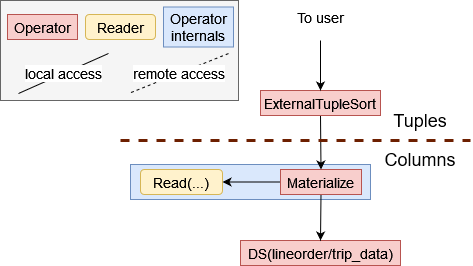}
    \caption{PosDB plans for queries used in experiments.}
    \label{fig:experiment1-plan}
\end{figure}

The PosDB query plan of the considered query is shown in Fig.~\ref{fig:experiment1-plan}. Essentially, it is equivalent to the PostgreSQL one. Despite the fact that PostgreSQL is a row-based system, whereas PosDB is column-based, the comparison is still valid since the query projects all attributes. As the result, PosDB behaves identically to the row-based system: readers of the Materialize operator access the same amount of data from disk, albeit from several different files. It is also worth mentioning that PostgreSQL uses a heap to merge runs. In all conducted experiments, the number of tapes was greater than the number of runs. This means that the polyphase merge algorithm behaved like a multiway merge both in PosDB and PostgreSQL.

\begin{figure}[h!]
\centering
\begin{minipage}{.49\textwidth}
  \centering
  \includegraphics[width=.99\linewidth]{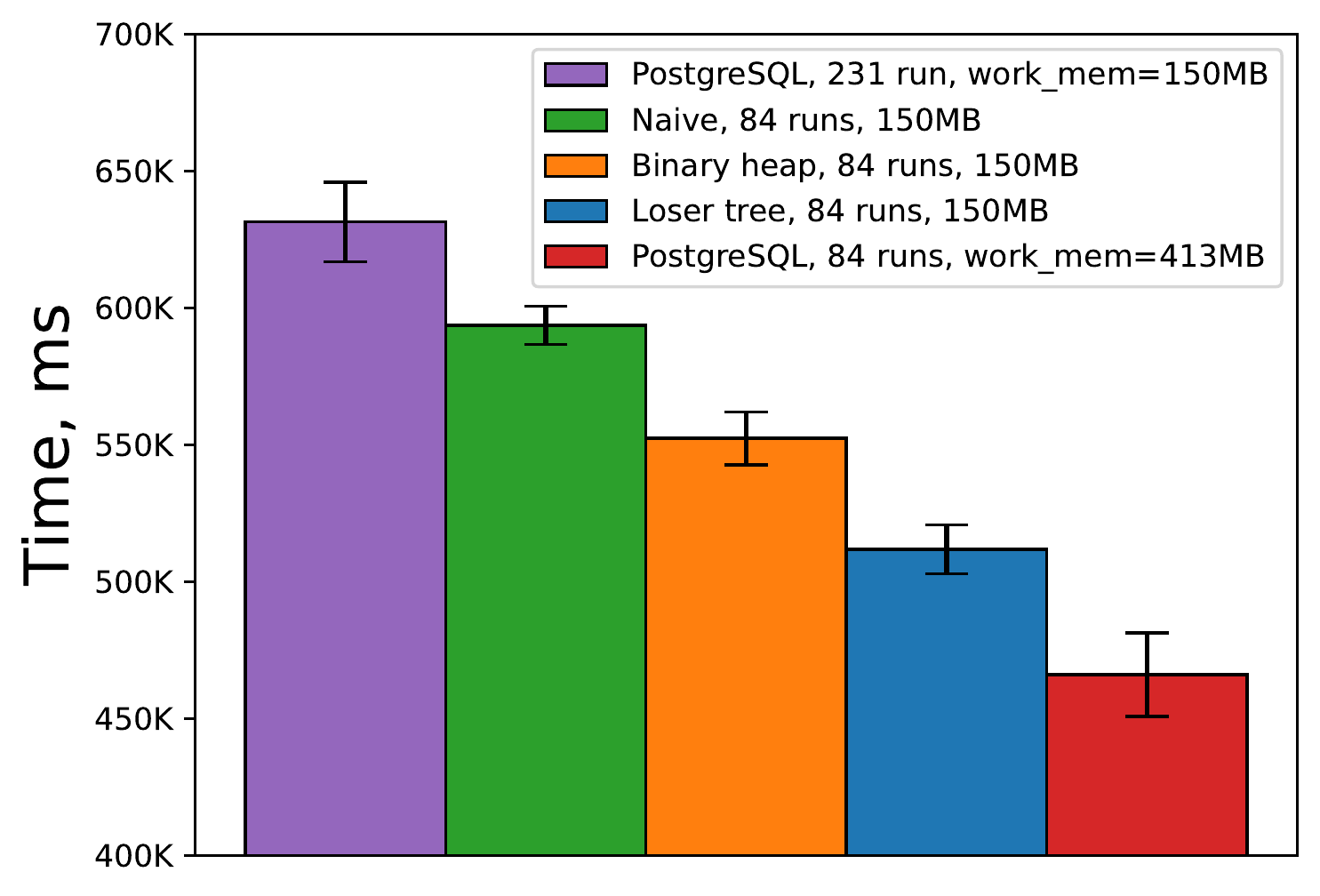}
  \captionof{figure}{PostgreSQL comparison (exp 1), evaluation on synthetic data (exp 2).}
  \label{fig:experiment-1}
\end{minipage}%
~~~~~
\begin{minipage}{.49\textwidth}
  \centering
  \includegraphics[width=.99\linewidth]{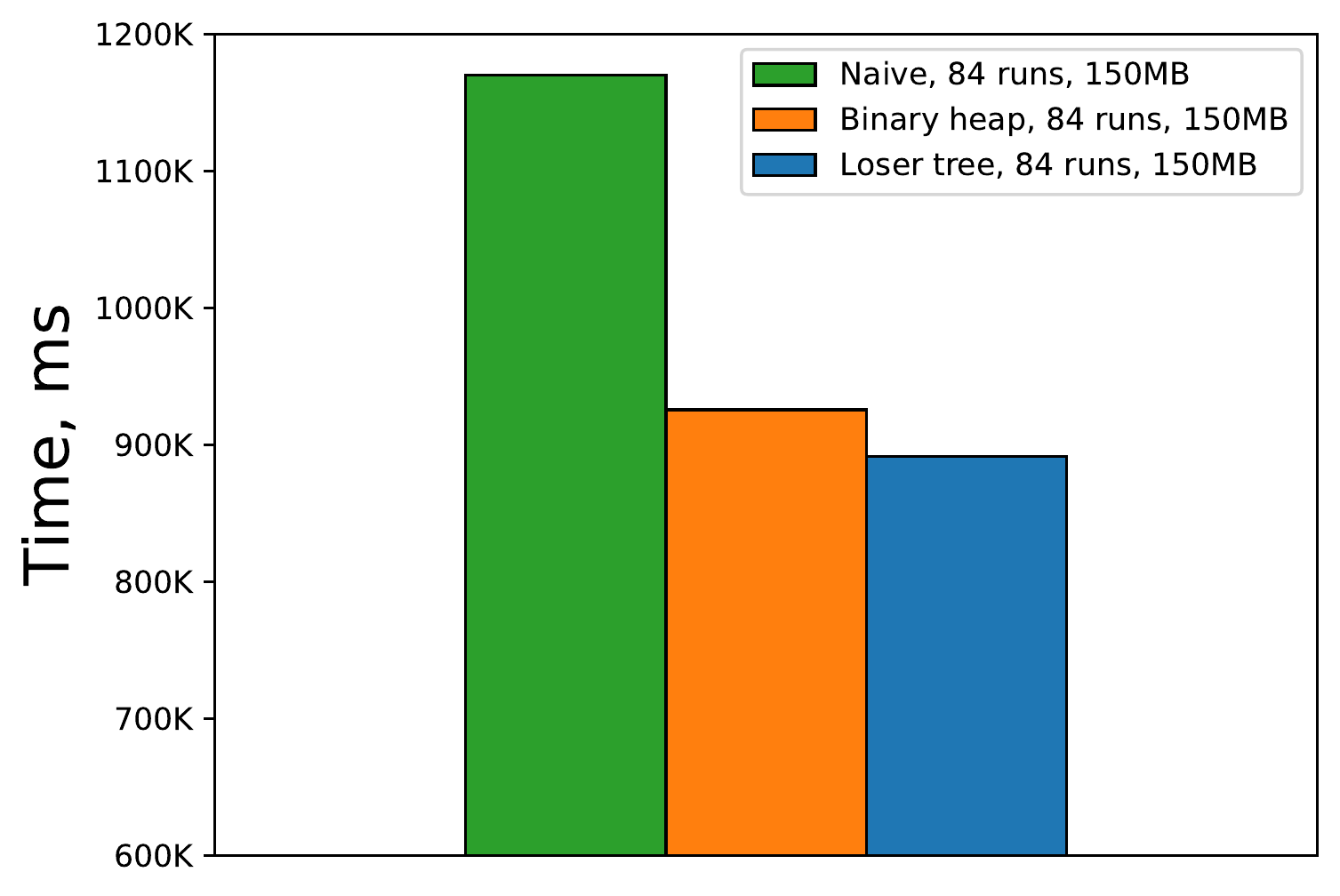}
  \captionof{figure}{evaluation on real data (exp 3), TripData dataset.}
  \label{fig:experiment-2}
\end{minipage}

\end{figure}

In the first two experiments, the total number of runs was chosen to be 84. PosDB generates 84 runs with 150 megabytes of memory available for the sort operator for all three considered algorithms. PostgreSQL at the same time needs 413 megabytes of \texttt{work\_mem}\footnote{\url{https://www.postgresql.org/docs/current/runtime-config-resource.html}} for this, probably because of internal buffers for optimized I/O. To make this comparison more accurate, we measured the performance of PostgreSQL with a total of 84 runs (\texttt{work\_mem}=413MB) and with 231 runs (\texttt{work\_mem}=150MB, the same amount of memory was given to the PosDB sort operator). In these experiments, the average query execution time for 40 iterations was taken with a confidence interval of 95\%. The results of the first experiment are shown in Fig.~\ref{fig:experiment-1}. PostgreSQL sorts 231 runs with 150MB of \texttt{work\_mem} a bit slower than the naive approach, but it sorts 84 runs considerably faster. There could be several reasons for this: PostgreSQL implements prereading for a better disk access pattern, plus it sorts \textlangle key prefix, pointer\textrangle\,pairs, which is more efficient~\cite{AlphaSort,GraefeSortingInDBMS} than pointer sorting.

Due to the space constraints, the results of the second experiment are shown in Fig.~\ref{fig:experiment-1} as well. It is easy to see that the naive approach is the worst, binary heap is the second, and loser tree is the best one out of the considered three.

In the third experiment we compared the performance of the implemented algorithms using the \texttt{trip\_data} table of the TripData dataset. Unlike the previous experiment, this one uses real data. The table has approximately 128 million rows (over 20 GBs). We used a simple query of the same kind as the one in the first experiment:

\begin{lstlisting}[language=sql, numberstyle=\color{mGray}, label={lst:query2}, numbers=left, basicstyle=\small, xleftmargin=2em,frame=single,framexleftmargin=1.5em]
SELECT medallion, hack_license, vendor_id, rate_code,
store_and_fwd_flag, pickup_datetime, dropoff_datetime,
passenger_count, trip_time_in_secs, trip_distance,
pickup_longitude, pickup_latitude, dropoff_longitude,
dropoff_latitude
FROM trip_data
ORDER BY trip_time_in_secs
\end{lstlisting}

The number of runs was set to 80. The query plan is similar to the one in the first experiment, and is also shown in Fig.~\ref{fig:experiment1-plan}.

The results of the third experiment are presented in Fig.~\ref{fig:experiment-2}. The naive approach is predictably much slower than the other two. Similarly to the previous experiments, loser tree shows better performance than binary heap.

Based on the obtained results, we can answer the research questions given in the abstract as follows: 1) using a specialized data structure at the merge step is beneficial and considerably speeds up sorting even in the case when disk is expected to be a bottleneck, 2) loser tree provides the best performance. 

\section{Threats to validity}

First of all, we would like to note that the behavior of the external sort operator (and specifically, the number of runs to merge) heavily depends on its available amount of memory. Performance of the merge step is determined by the number of performed comparisons, which is directly affected by the number of runs to merge. The less the number of runs to merge, the less noticeable will be the impact of a more efficient minimum element selection algorithm. For example, if the table to be sorted is two times larger than the memory available to the operator, then there will be two (or three) runs and speeding up the selection of minimal value will not bring any benefit.

We have considered a particular case when a DBMS sorts pointers to data values. It is stated~\cite{AlphaSort, GraefeSortingInDBMS} that sorting \textlangle key prefix, pointer\textrangle\,pairs is generally more efficient and speeds up the comparisons themselves, which leads to an increase in the intensity of disk I/O operations. Thus, total performance gains obtained from a specialized data structure may not be as significant. The same may be true for other advanced record representation and comparison techniques.

Next, in our study we consider SSDs and not HDDs. For a DBMS running on HDD, the ratio of CPU and disk expenses will be different since modern HDDs are much slower than SSDs. Therefore, the obtained speedup of the merge step may be lost to slow disk reads. In this case, reducing the number of comparisons via implementing a specialized data structure may not improve the overall performance that much.

In this paper, we have considered a single-threaded version of external sort. In the case of multi-threaded operator, the validation of our results would require a separate (and extensive) study since: 1) we would need to take into account the behavior of our data structure in case of multiple threads (e.g., cache pollution may happen), 2) disk could become the bottleneck and thus no prospective CPU gains would improve the overall performance. 

Finally, we focused on commodity PCs, and modern high-end servers would require a separate study as well.

\section{Acknowledgements}
We would like to thank Anna Smirnova for her help with the preparation of this paper.

\section{Conclusion}\label{sec:conc}
In this paper, we have considered the design of a comparison-based sort and evaluated the choice of the data structure used on the merge step for implementing the external sort operator in DBMSes. First, in our review of the related work, we demonstrated that no study has yet touched on the problem of selecting a data structure for run merging in external sort for DBMSes. After that, we posed several related research questions. To answer them, we have implemented three different approaches and experimentally evaluated them using the PosDB query engine. As the result, we have shown that a loser tree is the most efficient method for run merging.

Thus, we have demonstrated that even without complex I/O optimizations, such as disk striping, parallel I/O using \texttt{mmap}, elegant read ahead techniques, etc., the algorithm that selects the smallest element at the merge step has a noticeable impact on the performance of external sort. Therefore, it is worthwhile to make an effort to implement an efficient data structure for run merging on modern commodity computers. With the mentioned optimizations applied, the influence of the algorithm will only increase, since I/O operations become more efficient, which means there will be more computational load on the CPU. 

\bibliographystyle{splncs04}
\bibliography{bibliography.bib}

\begin{thebibliography}{10}
\providecommand{\url}[1]{\texttt{#1}}
\providecommand{\urlprefix}{URL }
\providecommand{\doi}[1]{https://doi.org/#1}

\bibitem{Abadi:2013:DIM:2602024}
Abadi, D., Boncz, P., Harizopoulos, S.: {The Design and Implementation of
  Modern Column-Oriented Database Systems}. Now Publishers Inc., Hanover, MA,
  USA (2013)

\bibitem{SortingOnNetworks}
Arpaci-Dusseau, A.C., Arpaci-Dusseau, R.H., Culler, D.E., Hellerstein, J.M.,
  Patterson, D.A.: High-performance sorting on networks of workstations. SIGMOD
  Rec.  \textbf{26}(2),  243–254 (jun 1997). \doi{10.1145/253262.253322},
  \url{https://doi.org/10.1145/253262.253322}

\bibitem{Chernishev:2018:PAO}
Chernishev, G.A., Galaktionov, V.A., Grigorev, V.D., Klyuchikov, E.S., Smirnov,
  K.K.: {{PosDB}}: {{An Architecture Overview}}. Programming and Computer
  Software  \textbf{44}(1),  62--74 (Jan 2018). \doi{10.1134/S0361768818010024}

\bibitem{DBLP:conf/dolap/ChernishevGGK022}
Chernishev, G.A., Galaktionov, V., Grigorev, V.V., Klyuchikov, E., Smirnov, K.:
  A comprehensive study of late materialization strategies for a disk-based
  column-store. In: Stefanidis, K., Golab, L. (eds.) Proceedings of the 24th
  International Workshop on Design, Optimization, Languages and Analytical
  Processing of Big Data {(DOLAP)} co-located with the 25th International
  Conference on Extending Database Technology and the 25th International
  Conference on Database Theory {(EDBT/ICDT} 2022), Edinburgh, UK, March 29,
  2022. {CEUR} Workshop Proceedings, vol.~3130, pp. 21--30. CEUR-WS.org (2022),
  \url{http://ceur-ws.org/Vol-3130/paper3.pdf}

\bibitem{DeWittSplitting}
DeWitt, D.J., Naughton, J.F., Schneider, D.A.: Parallel sorting on a
  shared-nothing architecture using probabilistic splitting. In: Proceedings of
  the First International Conference on Parallel and Distributed Information
  Systems. p. 280–291. PDIS '91, IEEE Computer Society Press, Washington, DC,
  USA (1991)

\bibitem{Samplesort}
Frazer, W.D., McKellar, A.C.: Samplesort: A sampling approach to minimal
  storage tree sorting. J. ACM  \textbf{17}(3),  496–507 (jul 1970).
  \doi{10.1145/321592.321600}, \url{https://doi.org/10.1145/321592.321600}

\bibitem{GPUTeraSort}
Govindaraju, N., Gray, J., Kumar, R., Manocha, D.: Gputerasort: High
  performance graphics co-processor sorting for large database management. In:
  Proceedings of the 2006 ACM SIGMOD International Conference on Management of
  Data. p. 325–336. SIGMOD '06, Association for Computing Machinery, New
  York, NY, USA (2006). \doi{10.1145/1142473.1142511},
  \url{https://doi.org/10.1145/1142473.1142511}

\bibitem{GraefeVolcano}
Graefe, G.: Query evaluation techniques for large databases. ACM Comput. Surv.
  \textbf{25}(2),  73–169 (jun 1993). \doi{10.1145/152610.152611},
  \url{https://doi.org/10.1145/152610.152611}

\bibitem{GraefeSortingInDBMS}
Graefe, G.: Implementing sorting in database systems. ACM Comput. Surv.
  \textbf{38}(3),  10–es (Sep 2006). \doi{10.1145/1132960.1132964},
  \url{https://doi.org/10.1145/1132960.1132964}

\bibitem{Hellerstein}
Hellerstein, J.M., Stonebraker, M., Hamilton, J.: Architecture of a database
  system. Found. Trends Databases  \textbf{1}(2),  141–259 (Feb 2007).
  \doi{10.1561/1900000002}, \url{https://doi.org/10.1561/1900000002}

\bibitem{ExactSplitting}
Iyer, B.R., Ricard, G.R., Varman, P.J.: Percentile finding algorithm for
  multiple sorted runs. In: Proceedings of the 15th International Conference on
  Very Large Data Bases. p. 135–144. VLDB '89, Morgan Kaufmann Publishers
  Inc., San Francisco, CA, USA (1989)

\bibitem{10.1145/3448016.3457319}
Jin, W., Qian, W., Zhou, A.: Efficient string sort with multi-character
  encoding and adaptive sampling. In: Proceedings of the 2021 International
  Conference on Management of Data. p. 872–884. SIGMOD '21, Association for
  Computing Machinery, New York, NY, USA (2021). \doi{10.1145/3448016.3457319},
  \url{https://doi.org/10.1145/3448016.3457319}

\bibitem{KnuthSort}
Knuth, D.E.: The Art of Computer Programming, Volume 3: (2nd Ed.) Sorting and
  Searching. Addison Wesley Longman Publishing Co., Inc., USA (1998)

\bibitem{LoadBalancedSorting}
Kumar, A., Lee, T.T., Tsotras, V.J.: A load-balanced parallel sorting algorithm
  for shared-nothing architectures. Distributed Parallel Databases
  \textbf{3}(1),  37--68 (1995). \doi{10.1007/BF01263656},
  \url{https://doi.org/10.1007/BF01263656}

\bibitem{Introsort}
Musser, D.R.: Introspective sorting and selection algorithms. Softw. Pract.
  Exper.  \textbf{27}(8),  983–993 (aug 1997)

\bibitem{AlphaSort}
Nyberg, C., Barclay, T., Cvetanovic, Z., Gray, J., Lomet, D.: Alphasort: A
  cache-sensitive parallel external sort. The VLDB Journal  \textbf{4}(4),
  603–628 (Oct 1995)

\bibitem{SSB}
O'Neil, P., O'Neil, E., Chen, X., Revilak, S.: The star schema benchmark and
  augmented fact table indexing. In: Nambiar, R., Poess, M. (eds.) Performance
  Evaluation and Benchmarking. pp. 237--252. Springer Berlin Heidelberg,
  Berlin, Heidelberg (2009)

\bibitem{BlockOriented}
Padmanabhan, S., Malkemus, T., Jhingran, A., Agarwal, R.: Block oriented
  processing of relational database operations in modern computer
  architectures. In: Proceedings 17th International Conference on Data
  Engineering. pp. 567--574 (2001). \doi{10.1109/ICDE.2001.914871}

\bibitem{AdaptiveSort}
Pang, H., Carey, M.J., Livny, M.: Memory-adaptive external sorting. In:
  Proceedings of the 19th International Conference on Very Large Data Bases. p.
  618–629. VLDB '93, Morgan Kaufmann Publishers Inc., San Francisco, CA, USA
  (1993)

\bibitem{Leyenda}
Shi, Y., Li, Z.: Leyenda: An adaptive, hybrid sorting algorithm for large scale
  data with limited memory. CoRR  \textbf{abs/1909.08006} (2019),
  \url{http://arxiv.org/abs/1909.08006}

\bibitem{JoinIndex}
Valduriez, P.: Join indices. ACM Trans. Database Syst.  \textbf{12}(2),
  218–246 (jun 1987). \doi{10.1145/22952.22955},
  \url{https://doi.org/10.1145/22952.22955}

\bibitem{8638394}
Watkins, A., Green, O.: A fast and simple approach to merge and merge sort
  using wide vector instructions. In: 2018 IEEE/ACM 8th Workshop on Irregular
  Applications: Architectures and Algorithms (IA3). pp. 37--44 (2018).
  \doi{10.1109/IA3.2018.00012}

\bibitem{TripData}
Whong, C.: New york city taxi fare data 2013  (2013),
  \url{http://chriswhong.com/open-data/foil\_nyc\_taxi/}

\bibitem{FlashMemorySorting}
Wu, C.H., Huang, K.Y.: Data sorting in flash memory. ACM Trans. Storage
  \textbf{11}(2) (mar 2015). \doi{10.1145/2665067},
  \url{https://doi.org/10.1145/2665067}

\bibitem{8855628}
Yin, Z., Zhang, T., Müller, A., Liu, H., Wei, Y., Schmidt, B., Liu, W.:
  Efficient parallel sort on avx-512-based multi-core and many-core
  architectures. In: 2019 IEEE 21st International Conference on High
  Performance Computing and Communications; IEEE 17th International Conference
  on Smart City; IEEE 5th International Conference on Data Science and Systems
  (HPCC/SmartCity/DSS). pp. 168--176 (2019).
  \doi{10.1109/HPCC/SmartCity/DSS.2019.00038}

\bibitem{SpeedingUpExternalMergesort}
Zheng, L., Larson, P.r.: Speeding up external mergesort. IEEE Trans. on Knowl.
  and Data Eng.  \textbf{8}(2),  322–332 (apr 1996). \doi{10.1109/69.494169},
  \url{https://doi.org/10.1109/69.494169}

\bibitem{10.5555/1972515}
Özsu, M.T., Valduriez, P.: Principles of Distributed Database Systems.
  Springer Publishing Company, Incorporated, 3rd edn. (2011)

\end{thebibliography}

\end{document}